# The Effect of Structural Conformational Changes on Charge Transfer States in a Light-Harvesting Carotenoid-diaryl-Porphyrin-$C_{60}$ Molecular Triad

Marco Olguin,[1] Rajendra R. Zope,[2] Tunna Baruah[1,2,a]

[1]Computational Science Program, University of Texas at El Paso, El Paso, TX 79968, USA

[2]Department of Physics, The University of Texas at El Paso, El Paso, TX 79968, USA

[a]Author to whom correspondence should be addressed. Electronic mail: tbaruah@utep.edu

We present a detailed study of charge transfer (CT) excited states for a large number of structural conformations in a light-harvesting Carotenoid-diaryl-Porphyrin-$C_{60}$ ($CPC_{60}$) molecular triad. The molecular triad undergoes a photoinduced charge transfer process exhibiting a large excited state dipole moment, making it suitable for application to molecular-scale opto-electronic devices. An important consideration is that the conformational flexibility of the $CPC_{60}$ triad impacts its dynamics in solvents. Since experimentally measured dipole moments for the triad of ~110 Debye (D) and ~160 Debye strongly indicate a range in conformational variability for the triad in the excited state, studying the effect of conformational changes on the CT excited state energetics furthers the understanding of its charge transfer states. We have calculated the lowest CT excited state energies for a series of 14 triad conformers, where the structural conformations were generated by incrementally scanning a 360 degree torsional (dihedral) twist at the $C_{60}$-porhyrin linkage and the porphyrin-carotenoid linkage. Additionally, five different $CPC_{60}$ conformations were studied to determine the effect of pi-conjugation and particle-hole Coulombic attraction on the CT excitation energies. Our calculations show that structural conformational changes in the triad show a variation of ~0.4 eV in CT excited state energies in the gas-phase. The corresponding calculated excited state dipoles show a range of 88 D – 188 D.

# INTRODUCTION

The elaborate and efficient photochemical energy conversion processes exhibited by photosynthetic reaction centers has stimulated much research into the design and synthesis of supramolecular artificial reaction centers which can mimic most of the major aspects of photosynthetic solar energy conversion.[1-11] Natural photosynthetic reaction centers constitute remarkable molecular-level photovoltaic devices which utilize incident photons to initiate a complex series of electronic transitions to achieve a high-energy charge separated state.[5,7] The understanding of the fundamental photosynthetic electronic transition pathways, which include singlet-singlet energy transfer, triplet-triplet energy transfer, and photo-initiated electron transfer, has provided a firm platform for the development of chemical systems which duplicate such efficient energy conversion processes.[3,7,12-33] The molecular building blocks employed in many artificial reaction centers for the successful mimicry of photosynthetic energy conversion usually consist of organic/inorganic pigments covalently linked to electron donor and/or acceptor moieties[17,18,21,22,24,25,34-43].

To emulate the light-absorbing property of chlorophylls, many artificial reaction centers feature porphyrins, chlorophyll derivatives, and related cyclic tetrapyrrolic molecules as the primary chromophore and excited-state electron donor.[3,12,16-18,20,31,34,36,37,40,42,44-49] Modeling natural photosynthesis led to the inclusion of quinones as the electron acceptor among some of the earliest synthesized photosynthetic mimics.[19,26,27,29,50-52] Subsequently, fullerenes were found to possess ideal electron acceptor qualities in artificial photosynthetic systems due to their large electron affinity, large charge accumulation capacity, and a small reorganization energy upon electron transfer.[53-60] These considerations led to the synthesis of a $CPC_{60}$ molecular triad consisting of a diarylporphyrin (P) covalently linked to a carotenoid polyene (C) and a $C_{60}$ fullerene.[33]

In natural photosynthetic systems, a high quantum yield of the final charge separated state is achieved through a series of short-range, fast, and efficient electron transfer transitions.[5] This strategy was exploited in the design of the $CPC_{60}$ molecular photovoltaic triad by employing two electron donors and one acceptor, in which two sequential electron transfers lead to a long-lived charge separated state.[33] Since many dyad-based artificial reaction centers suffer from rapid charge recombination, the triad succeeded in retarding charge recombination by the addition of a secondary donor (carotenoid) molecule which allowed for an increased separation between the particle and hole states.[14,33] The large distance between the donor and acceptor components in the $CPC_{60}$ triad leads to weak electronic coupling and slows the charge recombination process.[14]

The synthesis and photochemistry of the $CPC_{60}$ molecular triad established that fullerenes can act as effective primary electron acceptors in multi-component systems larger than porphyrin-fullerene dyads.[33] The molecular triad generated long-lived charge-separated states with high quantum yields even at low temperature. A salient photochemical feature observed in the $CPC_{60}$ triad is that the $^+C^\bullet\text{-}P\text{-}C_{60}^{\bullet-}$ charge-separated state recombines to yield a carotenoid triplet state rather than the molecular ground state, similar to photosynthetic reaction centers.[14] For the $CPC_{60}$ triad, the charge transfer $^+C^\bullet\text{-}P\text{-}C_{60}^{\bullet-}$ transition yields a singlet radical pair state which evolves into a CT triplet radical pair. The excited state triplet radical pair recombines to yield a $^3C\text{-}P\text{-}C_{60}$ triplet carotenoid state.[14]

The main charge-transfer transition pathway between the ground-state and the final $^+CPC_{60}^-$ charge-separated state involves a local excitation on the porphyrin moiety, followed by electron transfer to the adjacent $C_{60}$ component. Next, the carotenoid transfers an electron to the positively charged porphyrin to yield the final charge-separated state. A recent study suggests that the photoinduced charge separation process of the triad is driven by correlated motion of electrons and nuclei.[61] The experimentally determined lifetime for the $^+CPC_{60}^-$ charge separated state is ~170 nanoseconds in a 2-methyltetrahydrofuran solution.[33] In 2003, Smirnov and co-workers employed a transient dc photocurrent technique to study transient dipoles formed upon excitation of the porphyrin chromophore in the $CPC_{60}$ triad.[62] The large magnitude (>150 D) of the experimentally determined dipole of the triad conforms to the particle-hole picture of the charge separated excited state, in which the hole state resides on the carotenoid component and the particle state is localized on the $C_{60}$ fullerene at a large particle-hole separation (>30 Å).

A previous DFT study by our group on the ground state properties of the triad compared DFT-optimized structures of a linear triad and an elbow-shaped triad.[63] It was determined that the linear triad is energetically more stable than its elbow-shaped conformer counterpart.[63] The ground-state $CPC_{60}$ structure consists of a pyrolle-$C_{60}$ linked to a diaryl-porphyrin, where the porphyrin moiety is perpendicular to the aryl rings. The aryl rings, in turn, are coplanar with the carotenoid component. The porphyrin is connected to the carotenoid by an amide linkage. A comparison of the total density of states (DOS) for the triad with the DOS projected onto three subunits (pyrolle-$C_{60}$, carotenoid with amide linkage, and diaryl-porphyrin) revealed that the hybridization of the molecular orbitals belonging to different components is negligible such that the orbitals involved in charge-transfer excited state transitions are mostly localized on the parent components.[63] Moreover, the triad absorption spectrum represents a nearly linear combination of the spectra of the components. In a separate study, the optical absorption spectrum of the triad was calculated using a time-dependent DFT (TD-DFT) formalism.[64] Again, the decomposition of the spectrum into optical densities corresponding to the isolated components clearly demonstrated that the total spectrum is very well approximated by the sum of the component spectra. The main features and shape of the TDDFT spectrum were in good agreement with experiment, where a small shift of approximately 0.3-0.4 eV between the calculated and observed peaks was attributed to solvent polarization in the experimental measurements.[64]

In another study, the P-ΔSCF excited state method was applied to the study of charge transfer excitations in the $CPC_{60}$ molecular triad.[65] The large particle-hole distance in the final $^+CPC_{60}^-$ charge separated state allows for an accurate estimate of the charge transfer excitation energy within the separated fragment limit, where the CT energy is determined from the carotenoid ionization energy (IP), the fullerene electron affinity (EA), and the fullerene-carotenoid Coulomb interaction (1/R) according to Mulliken's equation (IP-EA-1/R). The P-ΔSCF excited state method gives an excitation energy of 2.46 eV for the $^+CPC_{60}^-$ charge separated state in gas-phase, which is in good agreement with the point-charge estimate (IP-EA-1/R) of 2.5 eV.[65] The P-ΔSCF study of $CPC_{60}$ also showed that the polarization of a solvent, represented as a discrete lattice, may influence the charge transfer process by stabilizing the large dipole moment of the particle-hole state.[65] The importance of solvent polarization is also brought out by the experimental study of Gust et al.[33] in which charge separation was observed in benzonitrile and (2-methyl)-tetrahydrofuran but not observed in toluene. The effect of the solvent is significant as the experimental CT energy of the $^+CPC_{60}^-$ state is significantly smaller than the gas-phase calculated value.[25,65]

The magnitude of the excited state dipole moment for the triad is 153 D, where experimental estimates were made in deriving the dipole value.[62] The approximations entail a simplification of the molecular shape (ellipsoid), underestimation of the Coulomb attraction between the polarizable chromophores as well as the oppositely charged carotenoid-fullerene components, and conformational changes for the charge separated state in solution. In regard to molecular shape, an alternative evaluation consisting of a decomposition of the total dipole moment into solute and solvent polarization contributions gave an experimental estimate of 163 D, which corresponds to a particle separation of ~34 Å.[62] The polarization interaction between the carotenoid and porphyrin chromophores will have the effect of shortening the particle-hole distance, which may reduce the dipole magnitude but not necessarily induce structural changes. On the other hand, a Coulombic interaction between the negatively charged $C_{60}$ fullerene and the positively charged carotenoid tail in the excited state may induce significant structural changes which can move the carotenoid chain into a closer distance to the fullerene component. Experimental evidence for such a folded conformer comes from a study of the triad in micelle nanoreactors suspended in water, where contractions in the molecular volume of the triad were attributed to entropy changes arising from solvent movements and possible conformational changes upon photoinduced electron transfer in generating a dipole of ~110 D.[66] The conformational changes of the triad in water were studied by Cheung et al.[67] using classical molecular dynamics simulations. This study has brought out that the linear structure favored in the gas phase is one of the least populated conformational states, which shows that the triad may undergo significant conformational changes in solution.

In the present investigation, we extend our two previous DFT studies of the $CPC_{60}$ triad with a combined ground- and excited-state electronic structure study of various conformational configurations of the linear-shaped triad. Since entropy changes in the triad/solvent system control the excited state charge transfer process, it becomes important to study the excited state properties of several different triad conformations. Since our calculations are on the gas-phase triad, this study brings out the changes in the CT energy due to the conformational changes only, separate from the electronic polarization effects due to the polar solvents. In order to gain insight into the large differences in structural conformation (linear vs folded) and excited state dipole moment magnitude (~160 D vs ~110 D) observed for the triad, we have calculated CT energies for a set of 19 distinct triad conformations. For each conformer, the calculated lowest excited state transition corresponds to the lowest CT state. In order to study changes in the CT energy due strictly to conformational variations, single-point calculations were performed for the structures reported in the present study. To the best of our knowledge, rigorous all-electron calculations have not been carried out for a series of triad conformers, mainly due to the large computational expense involved in the ground- and excited-state calculations of systems as large as the $CPC_{60}$ triad (207 atoms).

**COMPUTATIONAL METHOD**

The calculations reported here were carried out using Density Functional Theory (DFT) as implemented in the NRLMOL code.[68-70] We employed the Perdew, Burke, and Ernzerhof (PBE) exchange-correlation energy functional within the generalized gradient approximation for all calculations reported here.[71,72] The calculations were performed at the all-electron level using a large Gaussian basis set specially optimized for the PBE functional used in this work.[70] The basis set for a given atom is contracted from the same set of primitive gaussians. The numbers of the primitive gaussians, s-type, p-type, and d-type contracted functions, along with the range of the exponents are given in Table I. This basis set resulted in a total of 6170 basis functions for the triad conformers studied here. All ground-state and excited-state calculations were performed using spin-polarized wavefunctions. The P-ΔSCF excited state DFT method[73,74] has been implemented in the NRLMOL code and used here to determine the energies of the charge transfer excited state transitions. To obtain the excitation energy, an electron from the HOMO is placed in the LUMO and the self-consistent problem is solved using the perturbative ΔSCF method. The energy of the triplet state is obtained if the two unpaired electrons in the HOMO and LUMO orbitals are of the same spin. However, if two unpaired electrons in the HOMO and LUMO orbitals have opposite spin, then such a state is a mixed state (a 50–50 mixture of pure singlet and triplet states) with an energy that is an average of the singlet and triplet set. The energy of the singlet state is calculated using the Ziegler-Rauk method by subtracting the triplet energy from two times the energy of the mixed state.[75] The P-ΔSCF method provides accurate estimates of the experimentally obtained charge transfer excited state energies for a set of 12 supramolecular Tetracyanoethylene (TCNE)-hydrocarbon dyads.[73] Previously calculated CT excitation energies for porphyrin-$C_{60}$ co-facial dyads are in excellent agreement with the range of experimental values reported in the literature for similar porphyrin-fullerene systems.[76] The method has also been applied to the study of charge transfer energetics in relation to varying geometrical orientation of the tetraphenyl-porphyrin/$C_{60}$ (TPP/$C_{60}$) and (zinc)tetraphenyl-porphyrin/$C_{60}$ (ZnTPP/$C_{60}$) supramolecular dyads.[77]

TABLE I. The numbers of s-, p-, and d-type contracted functions, number of primitive gaussians and the range of the gaussian exponents used for each atom.

| Atom | s-type | p-type | d-type | Primitives | Exponent Range |
|------|--------|--------|--------|------------|----------------|
| C    | 5      | 4      | 3      | 12         | $2.22 \times 10^4 - 0.077$ |
| H    | 4      | 3      | 1      | 6          | $7.78 \times 10 - 0.075$ |
| N    | 5      | 4      | 3      | 13         | $5.18 \times 10^4 - 0.25$ |
| O    | 5      | 4      | 3      | 13         | $6.12 \times 10^4 - 0.10$ |

**RESULTS AND DISCUSSION**

The importance of studying various structural conformations of the $CPC_{60}$ molecular triad is more pronounced for the excited state than the ground state. For the ground state, X-ray and NMR data show the all-trans configuration for the amide/carotenoid backbone to be the most stable.[50,78] However, for the excited state the formation of a radical cation on the carotenoid component in the charge separated state may facilitate rotation about the carbon-nitrogen amide bond to generate various cis-conformations in the carotenoid backbone.[79,80] Since the one-electron oxidation of a carotenoid may lead to bond length equalization between single and double bonds in the conjugated amide/carotenoid backbone, a reduction in the rotational barrier for bonds that are formally double bonds in the ground state allows for more conformational flexibility in the cationic carotenoid component.[79,80]

In the present study, we undertook a systematic structural search for conformers resulting from torsions using the DFT method. Two different dihedral segments from the $CPC_{60}$ molecular structure were selected, denoted as PF and CP, through which various conformer structures were generated by dihedral-angle rotations about each of the chosen 4-atom dihedral segments (shown in figures 1 and 3). The PF (porphyrin/fullerene) designation corresponds to triad configuration variations where the structural modifications originate from torsions about the '($C_{60}$-pyrrole)-porphyrin' linkage. Similarly, the CP (carotenoid/porphyrin) designation describes structural changes effectuated through torsions about the 'porphyrin-carotene' amide linkage. For each dihedral segment, 7 distinct triad conformations were generated by successively incrementing the corresponding dihedral angle of the linear triad by 45 degrees for a full 360 degree torsion scan.

In figure 1 we show a dihedral segment of the $CPC_{60}$ triad at the linkage between the carotenoid and porphyrin subunits. The dihedral segment was scanned for full 360 degree torsion in increments of 45 degrees. At each of the 7 dihedral steps, the excited state energies for the lowest CT state (HOMO to LUMO transition) was calculated to determine the effect of varying conformational degrees of freedom for the amide/carotenoid backbone on the CT excitation energy (shown in figure 2). The calculated CT excitation energies for the lowest lying CT state of the 7 triad conformations (denoted as CP for carotenoid/porphyrin linkage) are given in Table II. For each CP conformation, we also report the calculated ground- and excited-state dipole moment magnitudes. The calculated CT energies lie within a range of 2.44 eV - 2.50 eV. The calculated excited state dipoles, corresponding to the particle state localized on the fullerene and the hole state localized on the carotenoid, are large and lie close to experimentally reported dipole values.

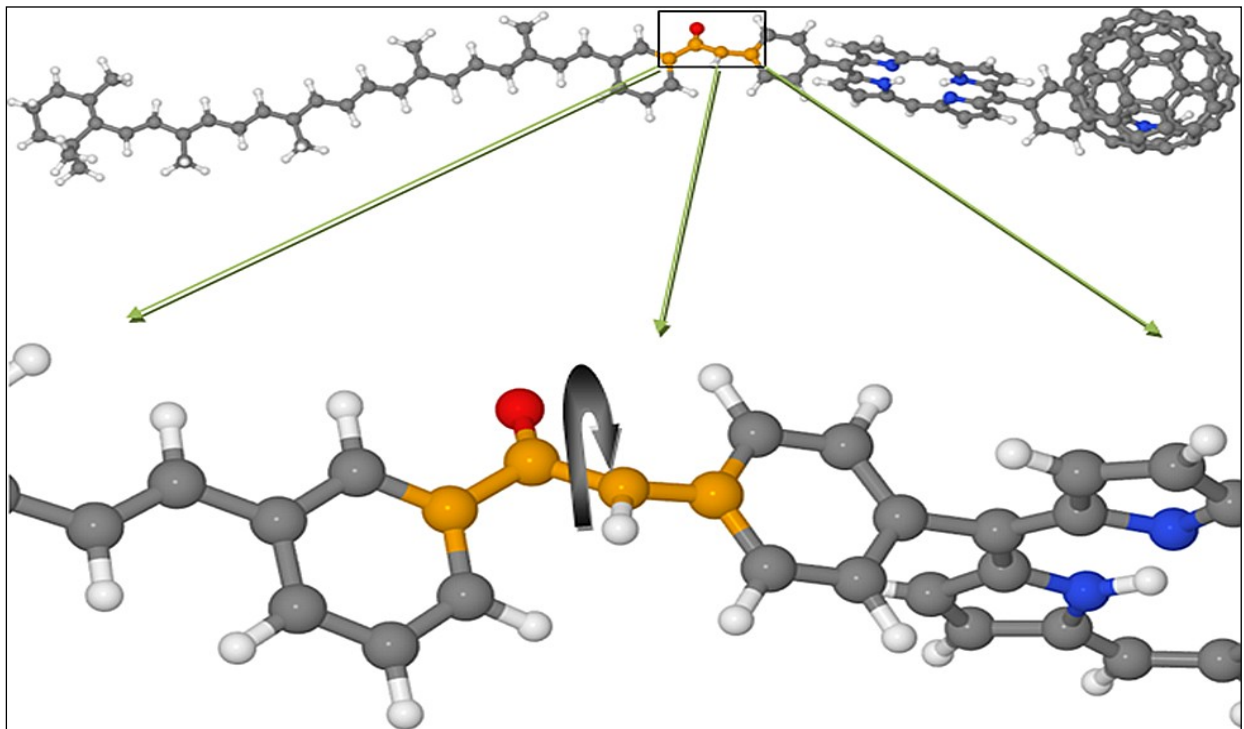

Figure 1. The 4-atom dihedral segment (colored orange) was used to generate a full 360° torsion scan consisting of 7 steps of 45° increments. The lowest CT excitation energies were calculated at each torsion-scan step.

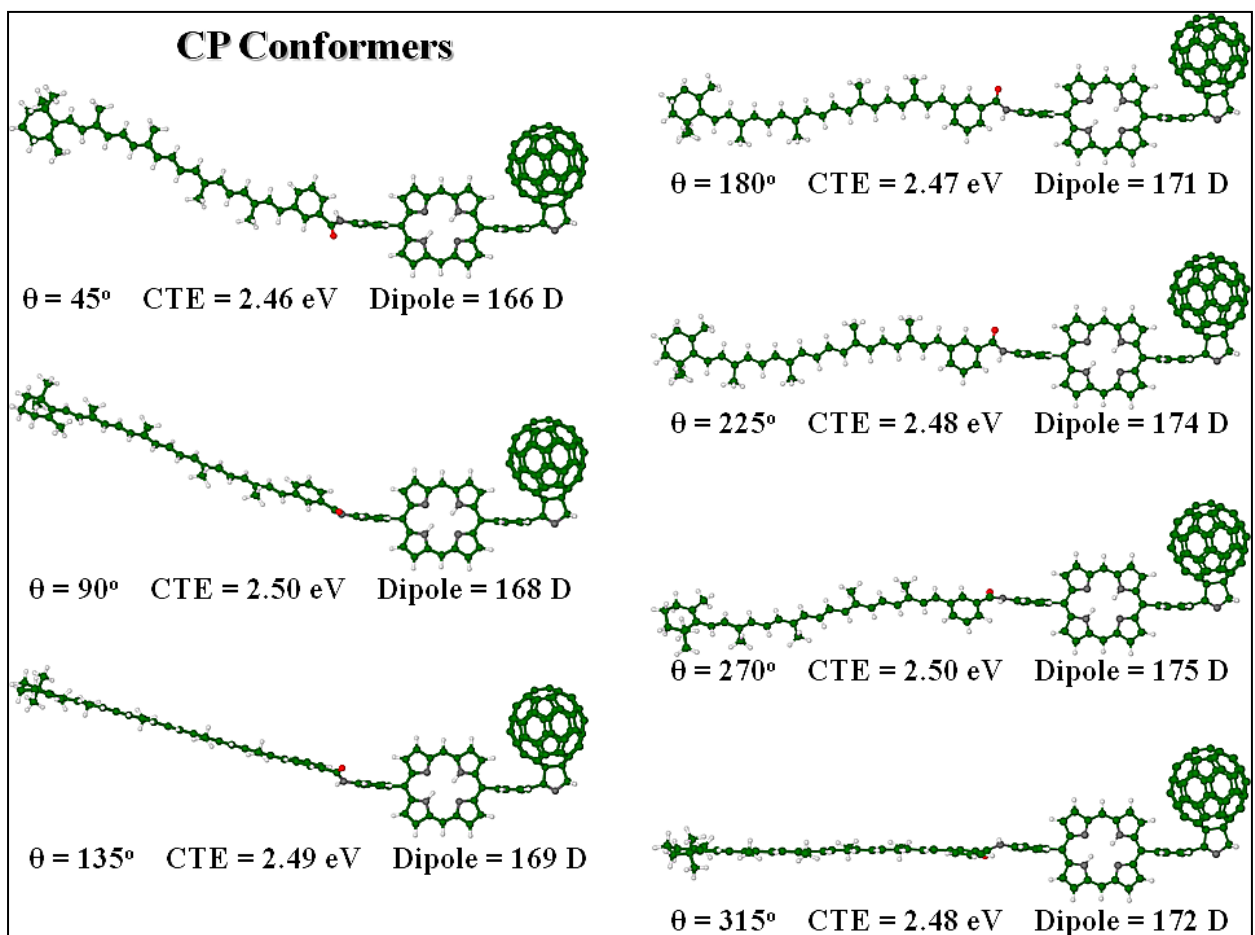

Figure 2. The angle (θ), charge transfer energy (CTE), and dipole magnitude values are shown for each of the 7 distinct CP conformers.

Table II. Charge transfer excitation energies (in eV) and ground- and excited-state dipole values (Debye) for the 7 triad conformations generated by torsions about the carotenoid/porphyrin linkage. CP denotes carotenoid/porphyrin.

| Triad Conformer | CT Excitation Energy | Ground State Dipole | Excited State Dipole |
|---|---|---|---|
| CP 45° | 2.46 | 7.9 | 165.6 |
| CP 90° | 2.50 | 9.2 | 168.0 |
| CP 135° | 2.49 | 9.0 | 169.4 |
| CP 180° | 2.47 | 8.0 | 171.1 |
| CP 225° | 2.48 | 8.4 | 173.7 |
| CP 270° | 2.50 | 9.3 | 174.6 |
| CP 315° | 2.48 | 8.6 | 171.7 |

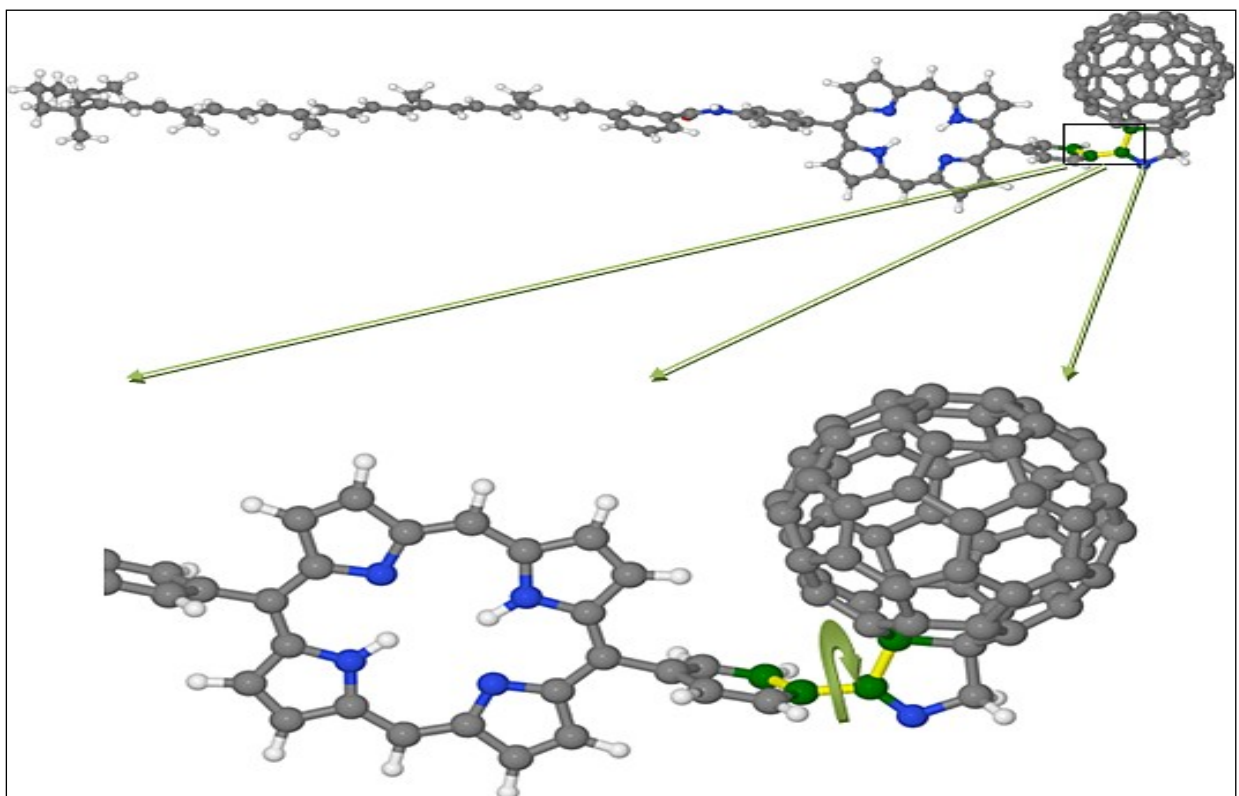

Figure 3. The 4-atom dihedral segment (colored green) was used to generate a full 360° torsion scan consisting of 7 steps of 45° increments. The lowest CT excitation energies were calculated at each torsion-scan step.

For the torsion displayed in figure 3 we have evaluated the CT energy at 7 dihedral steps in increments of 45 degrees for several low-lying CT states to determine the effect of varying conformational degrees of freedom for the porphyrin/$C_{60}$-pyrrole linkage on the CT excitation energy. The calculated CT excitation energies for the lowest lying CT state of the 7 triad conformations (denoted as PF for porphyrin/fullerene linkage) are given in figure 4 and Table III. For each PF conformation, we also report the calculated ground- and excited-state dipole moment magnitudes. The calculated CT energies lie within a range of 2.50 eV - 2.58 eV, which are slightly larger than the CT energies of the CP conformers. The calculated excited state dipoles for the PF conformers are larger than the CP conformers, where the values lie within the range of 170 D – 189 D.

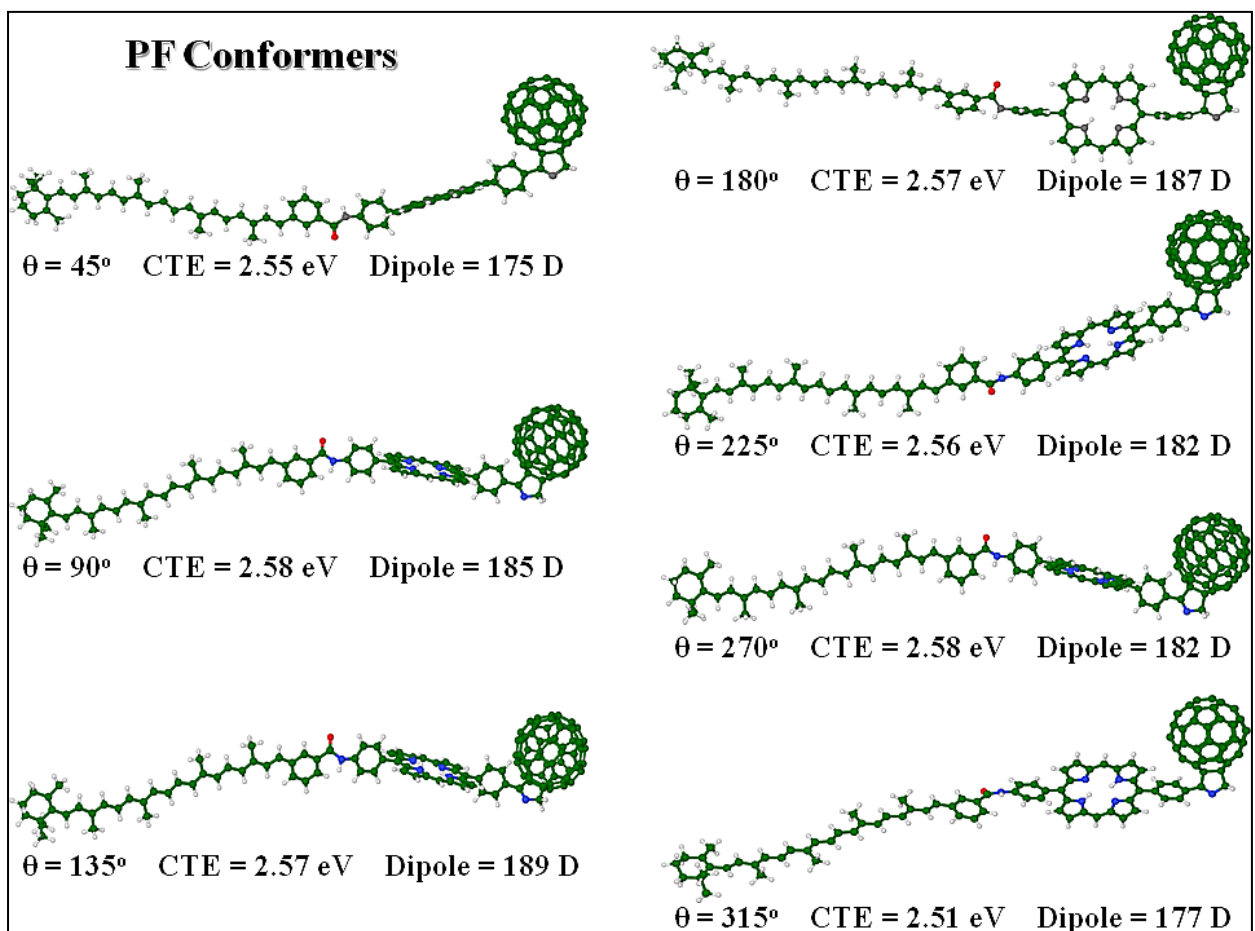

Figure 4. The angle (θ), charge transfer energy (CTE), and dipole magnitude values are shown for each of the 7 distinct PF conformers.

Table III. Charge transfer excitation energies (in eV) and ground- and excited-state dipole values (Debye) for the 7 triad conformations generated by torsions about the porphyrin/fullerene linkage. PF denotes porphyrin/fullerene.

| Triad Conformer | CT Excitation Energy | Ground State Dipole | Excited State Dipole |
|---|---|---|---|
| PF 45° | 2.55 | 8.8 | 174.9 |
| PF 90° | 2.58 | 10.6 | 185.4 |
| PF 135° | 2.57 | 10.5 | 188.7 |
| PF 180° | 2.57 | 9.7 | 187.2 |
| PF 225° | 2.56 | 9.4 | 181.8 |
| PF 270° | 2.58 | 10.6 | 182.1 |
| PF 315° | 2.51 | 9.5 | 176.5 |

The set of 14 triad conformers described as CP and PF above, was also optimized at the AM1, PM3, and PM6 semi-empirical levels of theory using the MOPAC2009[81,82] quantum chemistry software package where a similar energy ordering was obtained across the different semi-empirical methods. From the semi-empirical calculations of the 14 conformers and the linear triad, a set of 5 conformer structures with energies within a ~0.5 eV range of the linear-triad energy was selected for further geometry optimization using density functional theory with the NRLMOL code. The all-electron DFT optimization (6170 basis functions) of the five competing triad conformer structures exhibited a structural tendency toward the linear-shaped geometry of the triad. The only distinct structural feature shared by the five competing DFT-optimized conformers is a propensity toward the torsion of the porphyrin macrocylce plane with respect to a fixed $C_{60}$ and carotenoid. This is the same torsion examined in Table III. Thus, these torsions do not result in any significant change in the CT excitation energy or in a significantly different structure.

In the $CPC_{60}$ molecular triad, electron transfer is mediated by the covalent linkage joining the donor and acceptor components. The porphyrin component of the triad contains aryl rings as linkage groups at the meso-positions of the pi-conjugated macrocycle. The aryl linkage rings exhibit angles of 45°<Θ<90° with the porphyrin plane. Resonance stabilization serves as a major driving force for conformations which exhibit extended conjugation between the meso-aryl groups and the porphyrin moiety and a rotational motion about the single bond joining the aryl ring to the macrocycle may populate conformations with significant pi-pi overlap (as shown in figure 5). Consequently, extended conjugation between the pi-system of the aryl ring and the pi-system of the porphyrin macrocycle may influence the donor-acceptor electronic interaction, which in turn affects the electron-transfer rate.[13] Studies have addressed this issue by attempting to disrupt extended conjugative interactions by placing alkyl substituents at beta-pyrrolic positions.[83,84] The aim is to force strict orthogonality between the aryl molecular plane and the porphyrin macrocycle plane through steric repulsion effects of the alkyl substituents on the aryl ring. Several factors compete against an induced strict perpendicular aryl-porphyrin plane alignment, where the porphyrin macrocycle may undergo significant distortions due to the increased steric repulsion. In addition, conformations in solution will most likely sample a rotational motion about the single bond joining the aryl ring to the porphyrin macrocycle.[13,25]

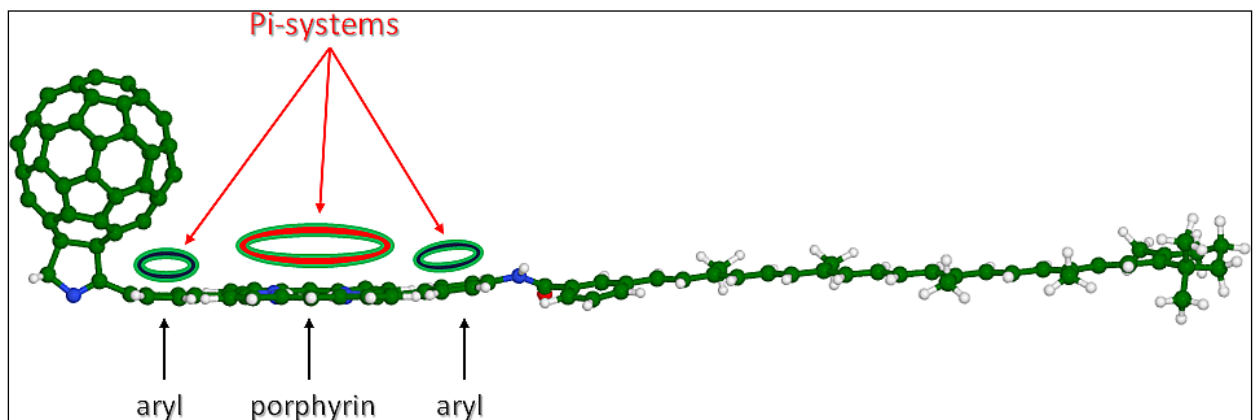

Figure 5. Simple Hückel type representations of the cyclic pi-conjugation exhibited by porphyrin and aryl systems.

In the present study, we examine the effect of extended conjugation between the pi-system of the aryl rings and the pi-system of the porphyrin macrocycle on the donor-acceptor CT excitation energies. We have defined a pseudo-dihedral parameter in figure 6. The dihedral parameter is evaluated, with respect to the ground-state dihedral, in 4 increments of 45 degrees for a total torsional scan of 180 degrees. At one end of the pseudo-dihedral scan, the pi overlap between the porphyrin macrocycle and its two aryl linkage groups at the meso-positions is maximal, whereas at the other end of the dihedral scan the extended pi-conjugation is completely disrupted by a relative perpendicular structural orientation between the porphyrin and aryl molecular planes. By calculating several low-lying charge transfer excitation energies at each dihedral increment, we can gain understanding into the effect of pi-conjugation coupling interactions on the CT excited-state energies. The calculated CT excitation energies for the lowest lying CT state of the 4 triad conformations are given in Table IV. For each pseudo-dihedral conformation, we also report the calculated ground- and excited-state dipole moment magnitudes. The calculated CT energies lie within a range of 2.48 eV - 2.53 eV and the calculated excited state dipoles lie within the range of 165 D – 176 D.

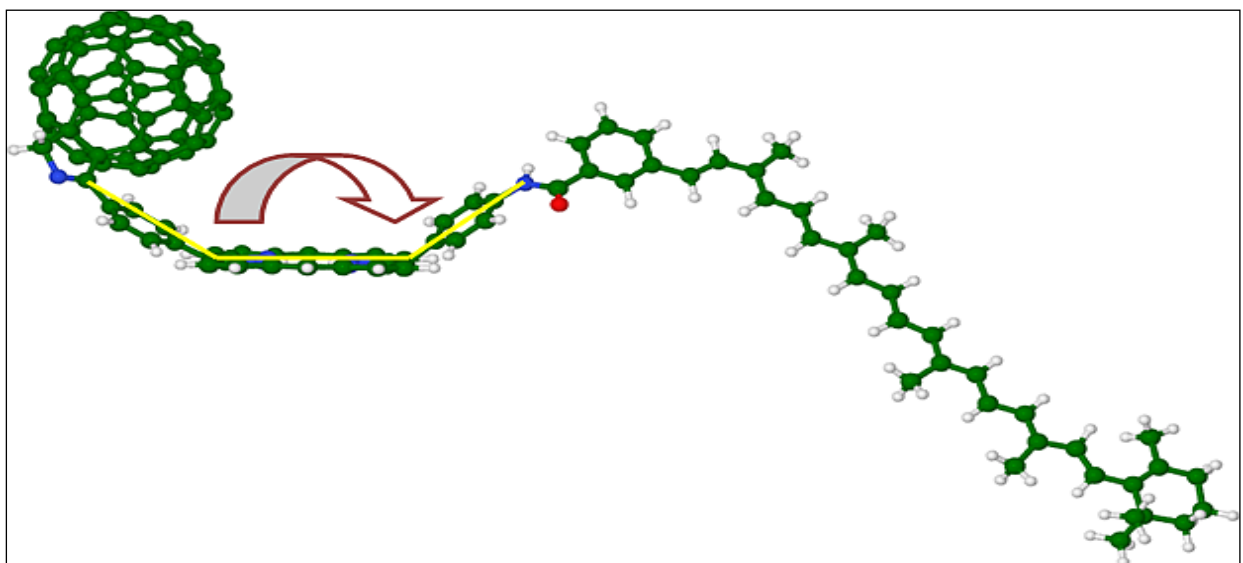

Figure 6. The pseudo-dihedral scan of 180° explores a varying degree of extended pi-conjugation between the porphyrin macrocycle and its two meso-aryl linkage groups.

Table IV. Charge transfer excitation energies (in eV) and ground- and excited-state dipole values (Debye) for the 4 triad conformations generated by torsions about the pseudo-dihedral.

| Triad Conformer | CT Excitation Energy | Ground State Dipole | Excited State Dipole |
|---|---|---|---|
| Dihedral 45° | 2.48 | 8.4 | 167.5 |
| Dihedral 90° | 2.50 | 8.8 | 165.4 |
| Dihedral 135° | 2.53 | 9.4 | 166.6 |
| Dihedral 180° | 2.50 | 9.2 | 176.0 |

Since the excited state dipole moments of all the conformers show large dipole moments, we constructed a folded geometry which is likely to possess a significantly lower excited state dipole moment as shown by experiment (~110 D).[66] In the folded configuration, the porphyrin-carotene extension wraps closely around the fullerene component where the $C_{60}$-carotene particle-hole distance is ~19 Å (shown in figure 7). The magnitude of the dipole moment for the $^+CPC_{60}^-$ charge-separated state of such a folded conformer would be smaller, which may reduce the photovoltaic efficiency in comparison to its larger dipole moment conformational counterparts. Also, the solvent stabilization of the charge transfer state would be less in comparison to the stabilization resulting from a larger molecular dipole such as in the linear triad conformation. Our calculated value for the CT excitation energy of the folded conformation yields a value of 2.16 eV, which is ~0.4 eV lower than the average CT energy of the CP and PF conformers. The excited-state dipole moment is reduced by half for the folded conformation (88 Debye) in comparison to the average dipole value for the CP and PF conformers. The energy of the $C^+PC_{60}^-$ charge transfer excited state is reported to be 1.20 eV by Gust and co-workers[25], which is significantly different from the gas-phase calculated value of 2.46 eV for the linear conformer. From the above results we find that the conformational changes can reduce the energy of the CT state in the gas-phase by a few tenths of an electron-volt even for conformers where the excited state dipole is significantly smaller. The remaining difference is likely to be due to the electronic polarization effects due to the polar solvent molecules.

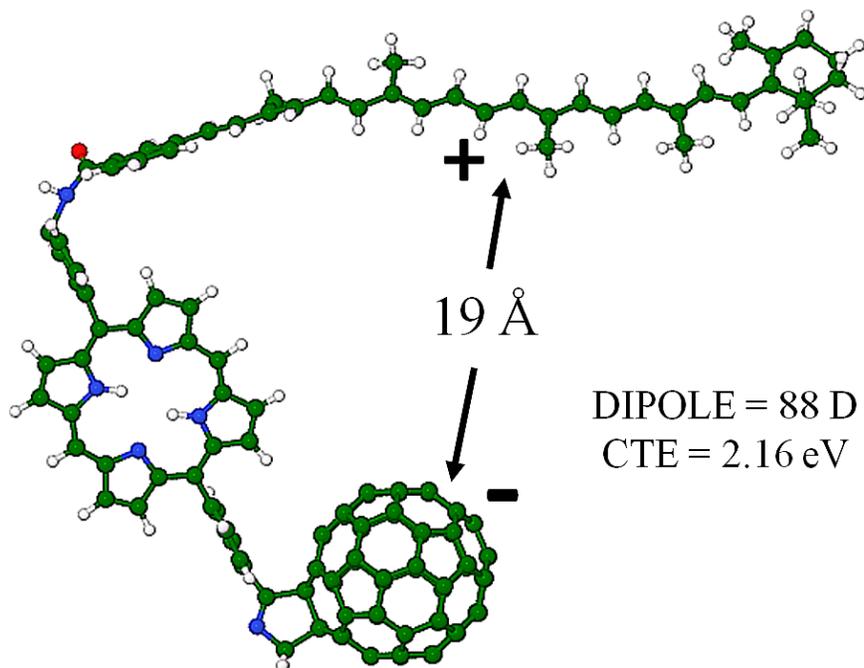

Figure 7. Folded triad conformation exhibiting a close interaction between the particle and hole states in the excited state (~19 Å). The Charge Transfer Energy (CTE) of 2.16 eV and the dipole moment value of 88 Debye are shown.

## CONCLUSIONS

We have examined the effect of structural conformational changes on the CT excited states of the $CPC_{60}$ molecular triad. Experimental evidence indicates that the triad undergoes conformational changes in solution, where the measured excited state dipole moments of 160 D and 110 D indicate a significant conformational variation in going from a linear structure to a folded conformation, respectively. To study the conformational flexibility of the triad, we generated a series of 14 distinct conformers through torsional scans about the carotene/porphyrin (CP) linkage and the porphyrin/fullerene (PF) linkage. In addition, we studied a folded conformation exhibiting a shorter donor-acceptor (carotene-fullerene) distance than the set of CP and PF triad conformers. Our calculations show that the CT excited state energy and the excited state dipole moment of the molecular $CPC_{60}$ triad varies slightly across the CP and PF triad conformers, where the calculated CT energy values for the triad lie close to 2.5 eV and the dipole values range from 165 D – 188 D. In comparison to the CP and PF conformers, the CT excited state energy of the folded conformation varies by ~0.4 eV with a value of 2.16 eV and the excited state dipole magnitude is reduced by half. Such a folded conformation will have strong implications on the CT excited state dynamics of the triad in solution since the reduced dipole magnitude will decrease the solvent-polarization induced stabilization of the CT state. Due to the structural flexibility of the $CPC_{60}$ triad, we also examined a conformational scan in which, at one end, the porphyrin macrocycle and its two meso-aryl groups lie nearly co-planar and at the other end, the molecular planes are perpendicular to each other. This particular conformational scan was designed to study the effect of extended pi-conjugation on the CT excited state energy of the triad. Although experimental studies show that such extended conjugation may affect the electronic coupling, which in turn impacts the CT rates, our calculations show that extended pi-conjugation exhibited by the porphyrin/aryl co-planar configuration does not produce a significant change in the CT excitation energy value. By studying the CT excited states of several triad conformations in the gas phase, we have de-coupled the effect of structural changes on the CT excitation energy from solvent effects on the CT energy. Since the calculated gas phase values for the CT energy differ from the reported experimental value, the solvent induced stabilization of the CT state becomes important. We are currently using a combined QM/MM approach to study the shift in CT excitation energy resulting from solvent stabilization.

## ACKNOWLEDGEMENTS

This work was funded by the Division of Chemical Sciences, Geosciences, and Biosciences, Office of Basic Energy Sciences of the U.S. Department of Energy (Grant No. DESC0002168). The support for computational time at the Texas Advanced Computing Center (TACC) from the National Science Foundation (NSF) (Grant No. TG-DMR090071) and from the National Energy Research Scientific Computing (NERSC) Center is gratefully acknowledged.